# A Markovian-based Approach For Daily Living Activities Recognition

Zaineb liouane[1], Tayeb Lemlouma[2], Philippe Roose[3], Fréderic Weis[4] and Hassani Messaoud[1]

[1]LARATSI – ENIM School, Monastir Tunisia, [2]IRISA – University of Rennes1, Lannion France, [3]LIUPPA/T2i – UPPA, Anglet France, [4]IRISA – University of Rennes1, Rennes France.
liouane.zaineb@yahoo.fr, Tayeb.Lemlouma@irisa.fr, Philippe.Roose@iutbayonne.univ-pau.fr, frederic.weis@irisa.fr and Hassani.Messaoud@enim.rnu.tn.

Abstract: Recognizing the activities of daily living plays an important role in healthcare. It is necessary to use an adapted model to simulate the human behavior in a domestic space to monitor the patient harmonically and to intervene in the necessary time. In this paper, we tackle this problem using the hierarchical hidden Markov model for representing and recognizing complex indoor activities. We propose a new grammar, called "Home By Room Activities Language", to facilitate the complexity of human scenarios and consider the abnormal activities.

## 1 INTRODUCTION

Elderly people have difficulties with notions of everyday life and are in a situation of dependency. They have difficulties or inability to perform redundant tasks as bathing, feeding; performs basic actions (getting up, moving); communicate (speaking, hearing).

The concept of smart home allows our seniors to continue to live as possible, while remaining free at home without changing their habits. Their own home is important as they have they habits and memories. Moreover, people live in a better health (physical & psychological) at home instead of being in old people's home as well as it is much more economic. This is why resources are currently organizing to keep our seniors at home. Thanks to new technologies, smart homes allow the notification of abnormal situations like fall, malaise or abnormal behavior and locate the person if (s) he is outside his/her home/garden. The recognition and the evaluation of human behavior, activities and interactions with the objects in a smart home is still an opened issue. In order to have the good, certain, accurate and realistic results, it is necessary to identify the efficient models and languages for the recognition of the behavior.

To describe the human behavior we propose a new grammar "Home By Room Activities language (HBRAL)" used to convert the real person scenario to a simple and comprehensive scenario. The principle of this grammar is to classify the activities of the person by room type in order to facilitate the recognition of normal and abnormal scenarios. Thereafter we will use the Hidden Markov model to predict the activities of elderly at home used our grammar. It is a well known framework to deal with uncertainty and dynamic data. Such model is often used for action recognition. In order to be more precise, we will use an extension called Hierarchical hidden Markov model *(HHMM)*; it is generally used for modeling complex activities in order to gain a precise recognition.

This model is a structured multi-level stochastic process that can be visualized as a tree structured variant of the *HMM*. It is suitable for the expression of user action data and it can find the prediction value about collected information and current situation. In this paper, we use the HHMM algorithm to obtain a reliable recognition person's activities. Thereafter, we calculate the likelihood ratio of the *HHMM* model between the prediction and the real activities. This paper is structured as follows: the next section presents the related works. In Section 3, we define a model for the recognition of scenarios and behaviors. In Section 4, we present the results of the implementation and simulation of our model. Finally, we present the estimation of our model we finish with a conclusion.

## 2 RELATED WORK

The functional tasks in daily lives of old seniors are divided into two parts, *ADL's* and *IADL's* (Msahli et al., 2014) and (Lemlouma et al.,2013). The activities of daily living *(ADL)* are the basic tasks of everyday life, such as eating, bathing, dressing, walking, toileting, and transferring.

The Instrumental activities of daily living *(IADL's)* are the activities that people do since they are awake such as dressing homework, phone use, etc. In this part we study the related work of the language used to describe the ADL and IADL's of the elderly precisely the language used to predict scenario of the person.

A recognition language is used to define a set of scenario to recognize the behavior of the person. Many researchers propose languages to recognize the human behavior in a smart home. In (Neyatia et al), authors propose a specific language: Human Behavior Scenario Description Language (HBSDL) to simulate the human dependency in a domestic environment and to describe the scenario of the human behavior during a large period of time. In (Zhang et al., 2011), authors present an extended grammar system SCFG (Stochastic Context-Free Grammars) for complex visual event recognition. It is based on rule induction and multithread parsing. In (Aritoni et al., 2011), the authors define the Event Recognition Language (ERL). It is a generative language able to define most of the events in daily life and especially the one interested in surveillance applications.

## 3 PROPOSED MODEL

Our study focuses on a particular kind of the resident that is *elderly* in order to provide them with required help and assistance. The considered scenarios includes: the person's behavior, the interaction with the system and surrounding objects and consider the person's degree of dependency. These scenarios will consider the constraints and difficulties that can face the resident is his daily life.

We define a *scenario* as the set of activities performed by the elderly person. The considered actions are those performed towards the system, such as: Off, On, Alarm, Warning. It is necessary to use an efficient model to recognize the scenario of the elderly person.

The majority of previous works were based on the Markov model as a model for the recognition of old people's activities in a smart home. Unfortunately, these models focus on particular events. For instance, (Singla et al., 2008) and (Kang et al., 2010) focuses on the "preparing diner" activity. Seen the good results obtained with the Markov model used for the recognition of particular activities, we choose to use it for the recognition of the main activities achieved by the resident during a day to take a generic and more developed solution. In order to obtain a good result, we should focus on the accuracy and precision of information to intervene as early as possible in case of emergency.

### 3.1 The Hierarchical Markov Models

This paper tackles the problem of studying and recognizing human activities of daily living (ADL), which is an important research issue in building a pervasive and smart environment. In dealing with ADL, we argue that it is beneficial to exploit both the inherent hierarchical organization of the activities and their typical duration. The Hierarchical Hidden Markov Model (HHMM) is an extension of the hidden Markov model to include a hierarchy of the hidden states for the recognition of complex actions. This model consists a layered structure of Markov Models (MM). On the top levels (the parent level) each state activates another MM on the child level. In this study we propose to use the HHMM, a rich stochastic model that has recently been extended to handle shared structures, for representing and recognizing a set of complex indoor activities.

The advantages of hierarchical recognition are: Recognition of various levels of abstraction, simplification of low-level models and response to novel data by decreasing details. In this paper, we apply the HHMM to predict and recognize the behavior of people in a smart home network.

### 3.2 The Grammar Proposition

In this section, we propose to use a grammar to recognize and simplify the complex activities; the aim of this grammar is to classify the structure of the person's activities and to give meaning of used model. The activity of daily living based on a set of reaction between information, object and environment in the "home". We can consider that each environment in the" home" has a specific activity. Indeed, the home environment includes the physical home structure and the place where the activities are achieved (number of rooms, type of rooms, etc.). For each room, we store data about: type (Kitchen/Living-room/ Bedroom/etc), width, length, height and a list of all the objects that exist in the room. One makes many activities. Each activity has its one environment in the "home". For example: "*the person cannot prepare a meal in the bedroom*", "*the person cannot sleep in the bathroom*", etc. In order to simplify the recognition of activities, we tailor each activity to the location where it is performed. We consider a common home architecture that consists of: Bedroom, Bathroom *(including toilets),* Kitchen and Living room. In this study, we propose to classify the activity of the person by the room of home as shown in table 1, we take consider the ADL and IADL activities.

The variables of the following algorithm are: **T.K:** Usual time passed in kitchen, **T.Bth :** Usual time passed in bathroom., **T.Bed:** Usual time passed in bedroom., **T.Lvr :** Usual time passed in Living Room. **New.TK:** New time passed in kitchen, **New.TBth:** New time passed in bathroom. **New.TBed:** New time passed in bedroom., **New.TLvr :** New time passed in Living Room. **Tic** and **toc**: predefined function to calculate the time spent in each state.

Our grammar "Home By Room Activities Language *(HBRAL)"* can be described using a hierarchical hidden Markov model (HHMM).

**Algorithm of HBRAL:** The following algorithm defines clearly the operation of our grammar:

```
Switch (type of room){
Case {Kitchen}
tic;
For (i=1;   i<=T.K;   i++)
{Activity=Activities-Kitchen;
Object=Object-Kitchen;}
New.T.K=toc;
break;
Case {Bathroom}
tic;
For (i=1;   i<=T.Bth;   i++)
{Activity=Activities-Bathroom;
Object=Object-Bathroom;}
New.T.Bth=toc;
break;
Case {Bedroom}
tic;
For (i=1;   i<=T.Bed;   i++)
{Activity=Activities-Bedroom;
Object=Object-Bedroom;}
New.T.Bed= toc;
break;
Case {Living room}
            tic;
    For (i=1;   i<=T.Lvr;   i++)
  {Activity=Activities-Livingroom;
  Object=Object-Living room;}
  New.T.Lvr = toc;
break;       }
```

Consequently we describe the HBARL using HHMM to obtain a better result.

In this paper, we apply the HHMM with a shared structure to predict and recognize the behaviors of the inhabitant in a smart home network for the elderly person.

The main and sub-activities are mapped into a shared-structure HHMM, which has fourth levels. Figure 1 shows the architecture of the HHMM model based on our grammar. We show the relationship between levels. These relationships are defined in a strict and mono-directional hierarchy: from top to down. Always the lowest level depends on the highest previous one. Each level contains the elements of the same nature starting from the root level (Level1). For example, the second level includes the set of possible places, the third level is for the objects of the home and the last level (level 4) represents all the actions and movements of the resident. Level 1 is the root environment. Level 2 is the main environment. Level 3 and 4 are the activities and objects, respectively. All these levels have a direct link between them, for example if the person turn-on the TV, the following action will be - most probably- watching TV. Consequently, we cannot pass from level *n* to level *n*+2. These links allow us to make a complete scenario starting with the kind of place where the action is realized by the resident. Concerning the "level 3"each activity can have from 1to n object(s) with n is the number of objects used in this activity. The link between each object in the same "level 4" is the <**and**>. Example: the person at home, in the kitchen, preparing meal, use stove **and** refrigerator.

Table 1: Activity and objects classed by the room.

|  | Kitchen | Bath room | Bed room | Living room |
|---|---|---|---|---|
| Activities | preparing a meal, eating, drinking, using a stove, washing | taking a shower, taking a bath, toileting | sleeping, dressing, reading a book, | Watching a TV, staying in a bank, read a journal book , drink a coffee, |
| Objects | Refrigerator Coffee filter Stove, dishwasher, etc. | Bath Sink Toilet, etc. | Clothes Radiator Bed, etc. | TV Radiator, etc. |

**Algorithm based on detection of abnormal activities:**

```
If ( New.T.K  >  PDT.K) then
   {Printf("alert      abnormal
activity kitchen");}
Else if( New.T.Bth  >   PDT.Bth)
then
   {Printf("alert      abnormal
activity bathroom");}
Else If ( New.T.Bed    >
PDT.Bed) then
   {Printf("alert      abnormal
activity bedroom");}
Else If ( New.T.Lvr    >
PDT.Lvr) then
   {Printf("alert      abnormal
activity living room");}end
```

**Detection of abnormal activities:** With our grammar HBRAL we can easily identify the abnormal activity precisely the unusual activities concerning the location in home. The last algorithm present the detection way of the abnormal activity. With **PDT**: Possible Delay Time.

$$PDT = \text{Usual time passed} + 30 \text{ minute}. \quad (1)$$

*Example:* possible Delay time in the kitchen
PDT.K = T.K+30.
We can notice that our algorithm helps the supervised to detect the abnormal activities related by the location.

**The advantage of the HBRAL described by HHMM:**
Our proposed model (HBRAL+HHMM**)** provides several advantage, first the principle of this model is to describe the person scenario from more general to more specific, The HBRAL based on HHMM reduce the ambiguity and the redundancy of data, additional  decrease the search filed and ensures easily the detection of abnormal concerned the location.

## 4 SIMULATION, RESULTS AND DISCUSSION

We propose to use the HBRAL grammar based on hierarchical hidden Markov model (Section 3.1) to recognize the activities of elderly. In this model, each level is simulated as a simple hidden Markov model following the normal law as a probability law with a standard deviation in order to evaluate the likelihood ratio of this model.

### 4.1 Implementation

We consider a piece of three rooms where each room is equipped with several sensors (presence, motion, etc.). The information provided by each sensor allows to know about the activities and the presence of the person. In our simulation, each rooms is linked to specific activity. The simulation period is five hours. We simulate these activities in a random way. Each scenario has a specific set of sensors regarding each room. In our case, we simulate these scenarios during *5* hours: from *7*:00 to *12*:00 A.M. Each room contains a specific scenario which contains from one to *n* activities. Example: at *7*:00 A.M. the person wakes up in the bedroom, takes toileting at *7:10* in the bathroom. In the kitchen, our resident prepares his lunch at *8:00*. Afterwards, he washes the dishes then go to the living room to watch TV at *9:15*. Later on, our resident has entered the kitchen to prepare his meal at *11*:00 then he takes his medication in the bedroom at *11:45*.

This example contains three scenarios: **The kitchen scenario:** preparing lunch at *8:00*; Wash these dishes at *8.30*; preparing meal at *11:00*. **The bedroom scenario:** wakes up at *7:00*; Take medication at *11.45*. **The living room scenario:** watch TV at *9.15*.

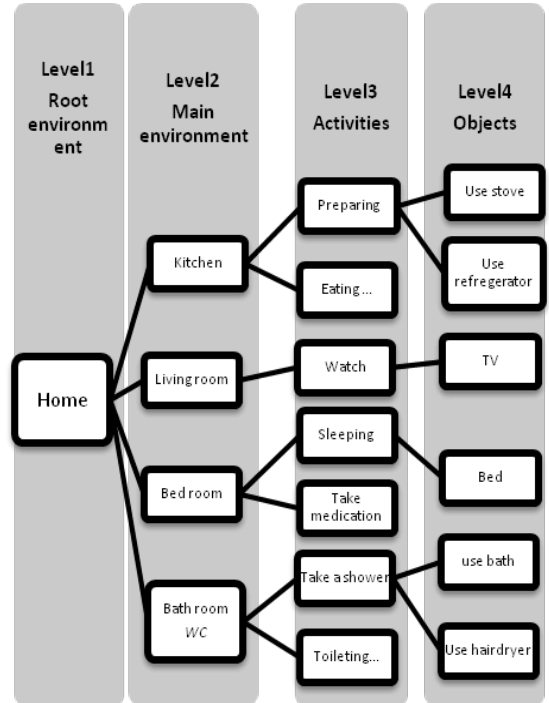

Figure 1: The architecture of the HHMM.

In order to simulate the current model, it is necessary to describe these parameters (N, A, B, *Π)* where N is the number of state, A is the transition matrix, B is the emission matrix and $\Pi$ is the initial matrix. The following matrices A, B and $\Pi$ are row stochastic, which means that each element is a probability and the elements of each row sum to 1, that is, each row is a probability distribution. In this case, we implement our scenarios with the following matrices.

*Transition matrix:* this matrix represents the probabilities of the transition between the states where each state represents a human scenario.

```
A              Scenaro K   Scenario L   Scenario B
Scenario K⎛     0.1         0.8          0.1    ⎞
Scenario L⎜     0.05        0.9          0.05   ⎟
Scenario B⎝     0.05        0.15         0.8    ⎠
```

*Initial matrix:* This matrix represents the probabilities of the initial state of our scenario.

```
       Scenario K    Scenario L    Scenario B
Π  (    0.7           0.2           0.1       )
```

*Emission matrix:* contains the emission probabilities, the probability to emit each observation for each state.

```
       Scenario K    Scenario L    Scenario B
B  (    0.1           0.7           0.2       )
```

## 4.2 Evaluations

We implemented the hidden Markov model using the Matlab. We tested performances of this model from the viewpoint of the recognition of behavior and human activity. Figure 2 shows the hidden states for each room (the first three curve) and the observation of our model (the fourth curve). Each targeted state has a specific scenario. The scenario kitchen is the activities realized in the kitchen according to exact times (the first curves), the scenario Living room is the activities realized in the kitchen according to exact times (the second curves), the scenario Bathroom is the activities realized in the kitchen according to exact times (the third curves). The fourth curves show the observation of these hidden states. Using the observation sequence $O$, the activities of the person can be estimated as shown in figure 2 using the transition matrix $A$ and the initial matrix $\Pi$. The fourth curve (*Observation*) shows the prediction of the three scenarios this curve present the evolution of the predicted hidden states that represent the evolution of the scenario observation over time; the observation $O_n$ is connected to the hidden state $Q_n$ at the same time.

$$O_n = h(Q_n) + V_n. \qquad (2)$$

With an additive noise $V_n$ independent $Q_n$.

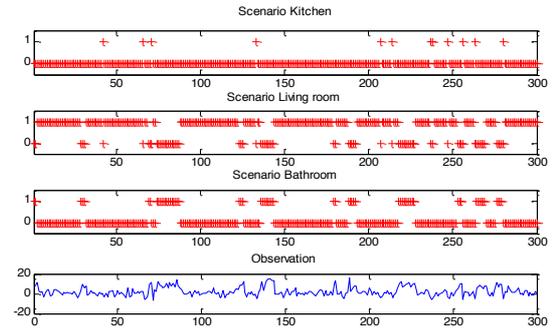

Figure 2: Hidden Markov Model: hidden state and observation.

Let $\Omega$ be a hidden Markov model and $O$ a sequence of acoustic observations. The recognition of this sequence is done by finding the $\Omega$ model that maximizes the probability $P(\Omega \mid O)$ (probability that $\Omega$ model generates a sequence of acoustic vectors $O$). This probability is also called *posterior probability*. Unfortunately, it is not possible to directly compute the $P(\Omega \mid O)$ probability. However, we can compute the probability that a particular model could generate a certain sequence of acoustic vectors $O$ i.e. $P(O \mid \Omega)$.

## 5 ESTIMATION

In this section we estimate the three scenarios *(Scenario L, Scenario K, Scenario B)* according to the time. Figure 3 shows the estimation of the hidden state" in this context the state is the place/location of the person" for each room in (*300 min*) using the scenario L (Living-room scenario), K (Kitchen) and B (Bath-room). Figure 3 shows the estimation of hidden states in terms of each room; each color represents the progress of each activity in terms of time. From Figure 3 we noticed that every activity runs independently.

### 5.1 Likelihood Measurements

To test the likelihood rate of obtained observation by the Markov model must compare our result "predicted" with the "real"; from this comparison we can see the performance and the accuracy of our detection model. From these results we can properly assess the errors to know the likelihood rate and we later conclude the efficiency of our results.

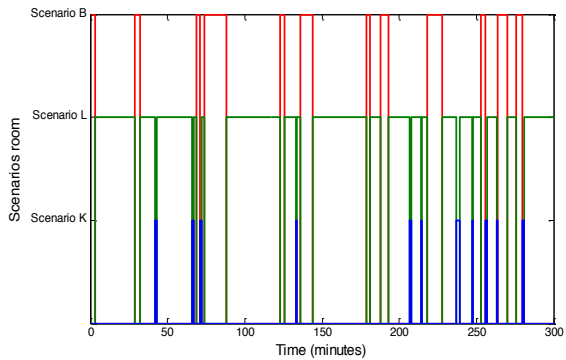

Figure 3: Hidden state estimation based on time.

## 5.2 Measures of forecast errors in the prediction of activities

We can obtain a prediction result that is very similar of our hidden state (real state). In this section, we estimate the error rate of our observation. ***Calculation of error***: Since the forecasts are usually false, a good forecast should also include a measure of predictable error. The error can be calculated using the difference between the prediction and the actual event: *Error = Actual event – Prediction*

$$E_t = R_t - P_t. \quad (3)$$

Figure 4 presents the error of observation in function of time. From Figure 4 we can see that the error of the prediction does not exceed the range [-1,1]. We noticed that for five hours the number of found errors is 9 which means that our model succeeds the prediction with only *3%* of error.

## 6 CONCLUSION

In this work, we were interested in the events recognition in smart environments for elderly and dependent persons. Our objective was to identify and experiment an efficient recognition model. We evaluated the likelihood rate of the hidden Markov model based on our grammar "Home By Room Activities language" in a smart home with a monitored person. Finally, we evaluated the efficient of this model using Matlab-based simulation tool. The results reveal that the proposed model is efficient for activities recognition with an observation error rate that is not very large compared to our hidden states. In the next steps of this work, we will explore the enrichment of our approach by investigating the learning-based systems (such as neural networks with a new learning algorithm) in order to recognize the events in a smart home and improve the learning phase such by using the *differential evolution* algorithm (Chengyao et al., 2015).

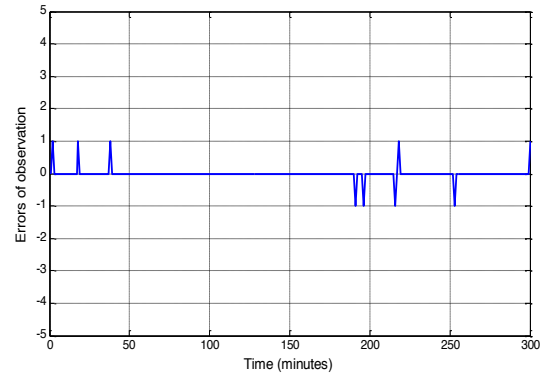

Figure 4: Error in function of time.